# Local Ridge Formation and Domain Delimitation in Aggregation-Diffusion Equations

Shin Nishihara [1*] and Toru Ohira [1]

[1] Graduate School of Mathematics, Nagoya University.

Furocho, Chikusaku, Nagoya, 464-8602, Japan.

*Corresponding author: shin.kurokawa.c8@math.nagoya-u.ac.jp, ORCID: 0009-0003-2351-0644

**Abstract**

On the carapace of turtles such as ***Mauremys japonica***, raised linear structures called keels form along the midline during embryonic development. This study investigates the underlying mechanisms of keel formation and domain delimitation on the carapace using a theoretical framework based on aggregation-diffusion equations. In this model, outward tissue growth is represented by density-dependent diffusion, while local ridge formation is modeled as distance-dependent aggregation potentially driven by haptotaxis. Analytical and numerical investigations reveal that distance sensitivity in aggregation plays a critical role in shaping ridge patterns and domain boundaries: low sensitivity promotes uniform density, whereas high sensitivity leads to localized high-density regions. The model may reproduce species-specific variations in keel formation, including the emergence of single or multiple keels, and accounts for the consistent appearance of the midline keel across diverse turtle species. Furthermore, the emergence of multiple high-density regions is shown to occur at points where the aggregation flux changes sign. These findings imply that cellular responses to structural gradients may underlie both ridge formation and boundary determination. This study helps possibly providing new insight into morphogenetic patterning on the turtle carapace and highlighting the role of distance-dependent cell aggregation in shaping complex biological structures.

## 1. Introduction

On the carapace of turtles such as *Mauremys japonica*, *Mauremys reevesii*, and *Trachemys scripta*, raised structures known as keels appear (Yasukawa et al., 2008; Okada et al., 2011; Akashi et al., 2022). Keel formation is observed during embryonic development, a stage in which the ribs grow outward from the midline (Okada et al., 2011; Paredes et al., 2020; Akashi et al., 2022). Through theoretical analysis of the morphogenetic mechanisms underlying keel formation, this study aims to provide insight - based on theoretical perspectives - into (i) the phenomenon of localized dorsal growth along the midline occurring during the phase in which the carapacial ridge (CR) promotes lateral rib growth (Cherepanov, 2006), and (ii) the still-unresolved functional role of the keel (Bang et al., 2016; Mayerl et al., 2018). Moreover, while a single keel along the midline appears in *M.japonica*, *M.reevesii*, and *T.scripta*, in *M.reevesii*, three keels (one along the midline and two lateral ones) are formed and maintained from the embryonic stage through to adulthood. In contrast, in *M.japonica*, the two lateral keels observed during embryogenesis disappear by adulthood, and in *T.scripta*, the lateral keels are absent during both embryonic and adult stages (Yasukawa et al., 2008; Okada et al., 2011; Moustakas-Verho et al., 2014; Akashi et al., 2022). In addition, observations of keel formation during embryogenesis in previous studies of *M.reevesii* and *Sternotherus odoratus* suggest that the single midline keel and the two lateral keels may differ morphologically (Paredes et al., 2020; Akashi et al., 2022). Taken together, these findings indicate that the two lateral keels exhibit variation among species and across developmental stages, whereas the midline keel consistently appears on the carapace. This suggests the existence of a conserved and species-independent developmental mechanism specifically governing midline keel formation. However, it should be noted that exceptions exist. For example, in *Platemys platycephala*, instead of a single keel forming along the midline, two keels are symmetrically positioned on either side of the midline. Thus, the formation of a single midline keel on the carapace cannot be generalized to all turtle species. Having said that, while all the species discussed so far are hard-shelled turtles, even in the softshell turtle *Apalone spinifera*, a single midline keel is observed (Greenbaum and Carr, 2002), suggesting that midline keel formation is a widely conserved feature across species. Another shared phenomenon is the establishment of the carapacial domain (boundary) during embryonic development. Although it is possible that the size of the ribs determines the carapace boundary, the mechanism by which rib growth is terminated remains unclear. Therefore, any attempt to elucidate a unified mechanism underlying keel formation must account for both the emergence of a single midline keel and the concurrent determination of the carapacial boundary during development.

To describe both of these phenomena, this study proposes an aggregation–diffusion equation as a theoretical framework. The outward growth of the carapace is represented by diffusion, while the formation of a keel along the midline is modeled as aggregation driven by haptotaxis (Murray, 2013). In particular, this study assumes that the diffusion coefficient in the diffusion term is density-dependent, reflecting the behavior of cells potentially involved in keel formation. It has been previously reported that when density pressure arises due to such density dependence, the system tends toward a uniform distribution of density (Shigesada et al., 1979; Mogilner and Edelstein-Keshet, 1999). This study examines the possibility and conditions under which spatial heterogeneity in density (*e.g.*, keel formation) can emerge even under density pressure, due to the presence of an aggregation term. It is also anticipated that, due to the nonlinearity of the aggregation–diffusion equation, analytical solutions may not always be obtainable. Therefore, for cases of unusual keel formation such as that observed in *P.platycephala*, where analytical solutions may be intractable, this study aims to reproduce the phenomena through numerical simulations. By verifying that the solutions obtained from the aggregation–diffusion equation successfully reproduce the two key features mentioned above, this study seeks to contribute to a deeper understanding of turtle carapace formation.

## 2. Modeling via Aggregation-Diffusion Equations

This study models the phenomenon in which individual cells on a two-dimensional domain both diffuse and aggregate. To facilitate understanding of the behavior on the two-dimensional domain, we first construct a model on a one-dimensional domain and then extend it to two dimensions.

### 2.1. Aggregation-Diffusion Equation on a One-Dimensional Domain

Following previous studies (Shigesada et al., 1979; Mogilner and Edelstein-Keshet, 1999), we assume that any point on the one-dimensional domain can be taken as the origin, and the diffusion coefficient is dependent on cell density. For the aggregation behavior, we assume, for instance, that based on cellular haptotaxis, components such as the extracellular matrix (ECM) are densely distributed at the origin, and cells aggregate in the direction of increasing ECM concentration (Murray, 2013). Specifically, the flux densities for aggregation and diffusion are expressed as follows:

$$J_A := -\varepsilon \left| \frac{x}{\chi} \right|^{\rho_1} \text{sgn}(x)\, u \,, \text{and } J_D := -\phi \left( \frac{u}{u_0} \right)^{\rho_2} \frac{\partial u}{\partial x}\,, \tag{1}$$

Here, $u[M]$ is a function of time $t[sec]$ and spatial variable $x[m]$, representing cell density. The parameters $\varepsilon[m/sec], \chi[m], \phi[m^2/sec],$ and $u_0[M]$ are positive constants, and $\rho_1$ is a non-negative constant. In particular, based on prior work, we assume $\rho_2 = 2$ (Shigesada et al., 1979), but to maintain generality, we continue to denote it simply as $\rho_2$. Additionally, as $x \to \infty$, we assume $u \to 0$, and $\partial u/\partial x \to 0$. The aggregation term indicates that aggregation speed varies depending on the distance from the origin. Furthermore, when $\rho_1$ is an odd number, the express can be simplified by removing the sign function and absolute value notation. Thus, the aggregation-diffusion equation on the one-dimensional domain is given as follows:

$$\frac{\partial u}{\partial t} + \frac{\partial}{\partial x} \left( J_A + J_D \right) = 0 \,. \tag{2}$$

Since the primary focus of this study is the resulting density distribution at equilibrium, we assume that a steady state is achieved as $t$ becomes sufficiently large, and seek $u(t, x)$ under this assumption. In other words, by solving:

$$J_A + J_D = 0 \,, \tag{3}$$

we obtain the equilibrium state $u(x)$ for sufficiently large $t$. The analytical solution is given by the following expression (see Appendix A):

$$u(x) = u_0 \left( \mathcal{C} - \frac{|x|^{\rho_1+1}}{\kappa(\rho_1+1)\chi^{\rho_1}} \right)^{\frac{1}{\rho_2}} \,,$$

$$\text{where } \mathcal{C} = \left( \frac{\kappa\chi^{\rho_1}}{(\rho_1+1)^{\rho_1}} \right)^{-\frac{\rho_2}{\rho_1+\rho_2+1}} \cdot \left( \frac{I}{\mathrm{B}\left(\frac{1}{\rho_2}+1, \frac{1}{\rho_1+1}\right)} \right)^{\frac{(\rho_1+1)\rho_2}{\rho_1+\rho_2+1}} \,, \text{and } \kappa := \frac{\phi}{\varepsilon\rho_2} \,. \tag{4}$$

Here, we assume that the total amount of cells remains constant over time (*i.e.*, $I := \int_0^\infty u(0, x)\mathrm{d}x$), and B denotes the beta function. As noted earlier, we assume $\rho_2 = 2$, which leads to the following boundary formation in the solution $u(x)$:

$$\left| x_{boundary} \right| = \left( \mathcal{C}\kappa(\rho_1+1)\chi^{\rho_1} \right)^{\frac{1}{\rho_1+1}} \,. \tag{5}$$

Furthermore, based on this analytical solution, it is evident that the highest density occurs at $x = 0$ within the formed domain. Utilizing this analytical solution of the aggregation-diffusion equation on a one-dimensional domain, we proceed to derive an analytical solution for the aggregation-diffusion equation on a two-dimensional domain, thereby investigating density inhomogeneity and domain formation.

## 2.2.  Aggregation-Diffusion Equation on a Two-Dimensional Domain

As in the previous section, we describe the aggregation-diffusion equation on a two-dimensional domain:

$$\frac{\partial u}{\partial t} + \frac{\partial}{\partial x}\left(J_{Ax} + J_{Dx}\right) + \frac{\partial}{\partial y}\left(J_{Ay} + J_{Dy}\right) = 0 \text{ , where}$$

$$J_{Ax} := -\varepsilon\left|\frac{x}{\chi}\right|^{\rho_{1x}}\text{sgn}(x)\,u \text{ , } J_{Dx} := -\phi\left(\frac{u}{u_0}\right)^{\rho_2}\frac{\partial u}{\partial x} \text{ ,}$$

$$J_{Ay} := -\varepsilon\left|\frac{y}{\chi}\right|^{\rho_{1y}}\text{sgn}(y)\,u \text{ , } J_{Dy} := -\phi\left(\frac{u}{u_0}\right)^{\rho_2}\frac{\partial u}{\partial y} \text{ .} \tag{6}$$

Here, $\rho_{1x}$, $\rho_{1y}$ are non-negative constants. Following the same assumption as in the one-dimensional case, we assume the existence of a steady state for sufficiently large $t$, and seek the solution $u(x,y)$. The analytical solution is given as follows (see Appendix B):

$$u(x,y) = u_0\left(\mathcal{C}_2 - \frac{|x|^{\rho_{1x}+1}}{K_x} - \frac{|y|^{\rho_{1y}+1}}{K_y}\right)^{\frac{1}{\rho_2}},$$

$$\text{where } K_x := \kappa(\rho_{1x}+1)\chi^{\rho_{1x}} \text{ , } K_y := \kappa(\rho_{1y}+1)\chi^{\rho_{1y}} \text{ ,}$$

$$\mathcal{C}_2 = \left[\frac{I_2(\rho_{1x}+1)(\rho_{1y}+1)}{4K_x^{\frac{1}{\rho_{1x}+1}}K_y^{\frac{1}{\rho_{1y}+1}}\,\text{B}\left(\frac{1}{\rho_2}+1,\frac{1}{\rho_{1y}+1}\right)\cdot\text{B}\left(\frac{1}{\rho_{1y}+1}+\frac{1}{\rho_2}+1,\frac{1}{\rho_{1x}+1}\right)}\right]^{\left(\frac{1}{\rho_{1x}+1}+\frac{1}{\rho_{1y}+1}+\frac{1}{\rho_2}\right)^{-1}}. \tag{7}$$

The total amount of cells is denoted by $I_2$. As in the one-dimensional case, since we assume $\rho_2 = 2$, the solution $u(x,y)$ develops the following boundary:

$$\mathcal{C}_2 - \frac{|x|^{\rho_{1x}+1}}{K_x} - \frac{|y|^{\rho_{1y}+1}}{K_y} = 0 \text{ .} \tag{8}$$

As in the previous section, this analytical solution shows that the highest density occurs at $(x,y) = (0,0)$ within the formed domain. We will later examine the specific shape of the boundary and whether density inhomogeneity exists within that domain. Furthermore, treating $\rho_{1x}, \rho_{1y}$ as variables, the area $S(\rho_{1x}, \rho_{1y})$ enclosed by the boundary can be expressed as follows (see Appendix C):

$$S(\rho_{1x}, \rho_{1y}) = \frac{4K_x^{\frac{1}{\rho_{1x}+1}}K_y^{\frac{1}{\rho_{1y}+1}}\mathcal{C}_2^{\frac{1}{\rho_{1x}+1}+\frac{1}{\rho_{1y}+1}}}{\rho_{1x}+1}\,\text{B}\left(\frac{1}{\rho_{1y}+1}+1,\frac{1}{\rho_{1x}+1}\right) \text{ .} \tag{9}$$

We will later investigate how this area (e.g., the size of the carapace) varies with changes in $\rho_{1x}, \rho_{1y}$.

As previously discussed, the analytical solutions of the aggregation-diffusion equations on both one-dimensional and two-dimensional domains show that the highest density occurs at $x = 0$, and $(x, y) = (0,0)$, respectively. These points correspond to where $J_A$, and $J_{Ax}, J_{Ay}$ change sign from positive to negative. In other words, if the aggregation flux is defined such that it changes from positive to negative at two locations, high-density regions are expected to form at those two points. For instance, consider the case where the aggregation flux is defined as follows:

$$J_{Ax} := -\varepsilon\alpha\left(\frac{x}{\chi}\right)\left(\left(\frac{x}{\chi}\right)^2 - \sigma^2\right) u \text{ , where } \alpha \text{ and } \sigma \text{ are positive constants,} \tag{10}$$

It is then expected that two high-density regions will form along the $x$-axis at $x = \pm\sigma$. However, since an analytical solution cannot be obtained for this case, numerical analysis is required to examine the behavior of the aggregation-diffusion equation under this flux.

## 3. Result

### 3.1. Validation of the Analytical Solution

First, we present an analytical solution of the aggregation-diffusion equation on a two-dimensional domain by assigning specific values to the parameters and illustrating the resulting solution. Since one of the main interests lies in whether the solution exhibits density inhomogeneity, we begin with a spatially uniform initial condition, with the total quantity normalized to $I_2$. As shown below, when $\rho_{1x} = \rho_{1y} = 3$, we confirm - consistent with previous studies - that the density remains uniform under the condition $\rho_2 = 2$ (Shigesada et al., 1979). Furthermore, it can also be confirmed that the domain has a finite area.

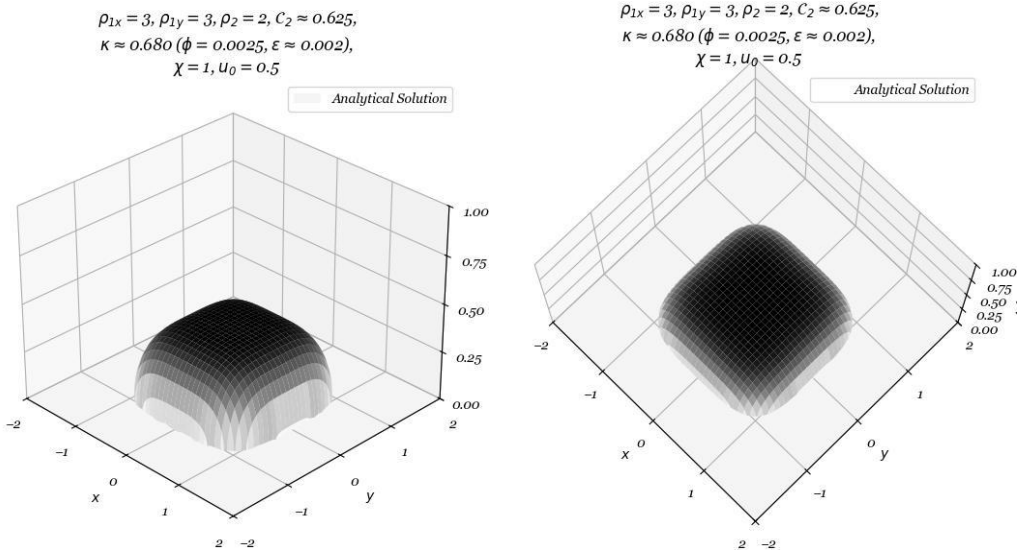

**Fig. 1**: Based on (7), the following parameters are set, and the resulting numerical analysis of $u(x, y)$ is presented in the graph: $\rho_{1x} = 3, \rho_{1y} = 3, \rho_2 = 2, \kappa \approx 0.680$ (based on $\phi = 0.0025, \varepsilon \approx 0.002$), $\chi = 1, u_0 = 0.5, I_2 = 3$, which leads to $\mathcal{C}_2 \approx 0.625$. The darker regions represent higher values of $u$, while the lighter regions represent lower values. The graph on the left is drawn from a horizontal perspective to show the uniformity or non-uniformity of the density $u$, while the graph on the right is drawn from a top-down view to show the boundaries of density $u$. These graphs are the same, with only the viewing angle changed.

On the other hand, when $\rho_{1x} = 0, 1$, the uniformity of the density breaks down even under the condition $\rho_2 = 2$, and a notable density inhomogeneity emerges particularly in the vicinity of the midline at $x = 0$.

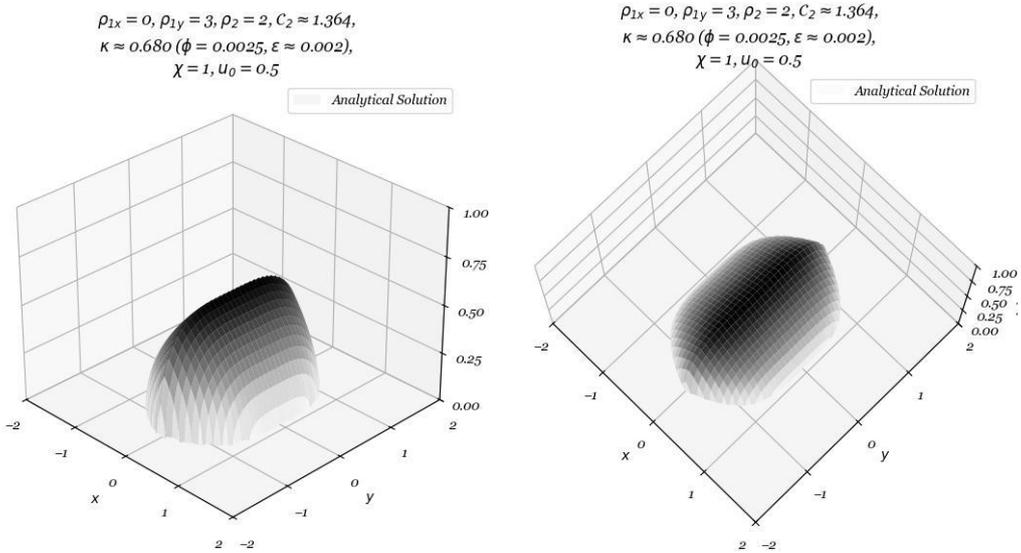

**Fig. 2**: Based on (7), the following parameters are set, and the resulting numerical analysis of $u(x, y)$ is presented in the graph: $\rho_{1x} = 0, \rho_{1y} = 3, \rho_2 = 2, \kappa \approx 0.680$ (based on $\phi = 0.0025, \varepsilon \approx 0.002$), $\chi = 1, u_0 = 0.5, I_2 = 3$, which leads to $\mathcal{C}_2 \approx$

1.364. The darker regions represent higher values of $u$, while the lighter regions represent lower values. The graph on the left is drawn from a horizontal perspective to show the uniformity or non-uniformity of the density $u$, while the graph on the right is drawn from a top-down view to show the boundaries of density $u$. These graphs are the same, with only the viewing angle changed.

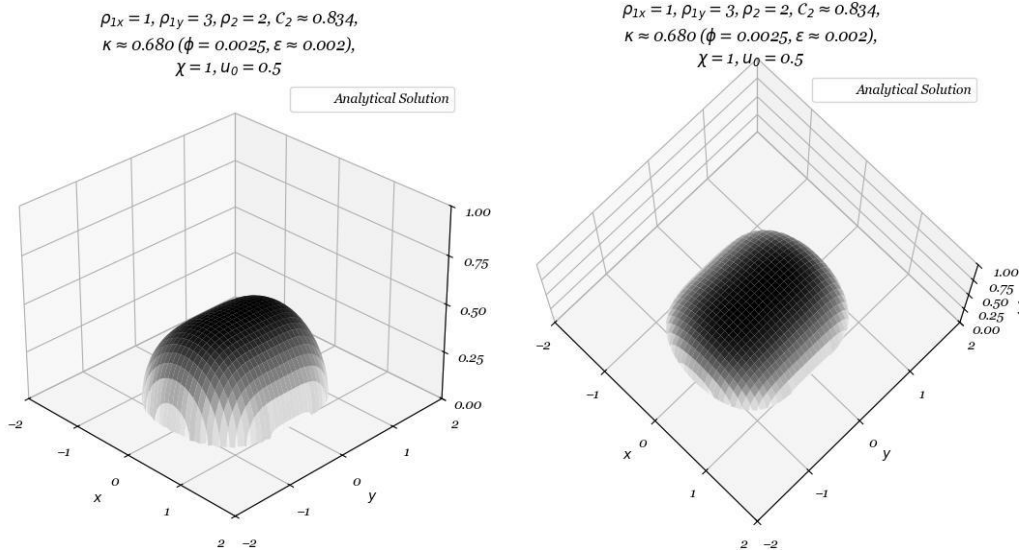

**Fig. 3**: Based on (7), the following parameters are set, and the resulting numerical analysis of $u(x, y)$ is presented in the graph: $\rho_{1x} = 1, \rho_{1y} = 3, \rho_2 = 2, \kappa \approx 0.680$ (based on $\phi = 0.0025, \varepsilon \approx 0.002$), $\chi = 1, u_0 = 0.5, l_2 = 3$, which leads to $\mathcal{C}_2 \approx 0.834$. The darker regions represent higher values of $u$, while the lighter regions represent lower values. The graph on the left is drawn from a horizontal perspective to show the uniformity or non-uniformity of the density $u$, while the graph on the right is drawn from a top-down view to show the boundaries of density $u$. These graphs are the same, with only the viewing angle changed.

In other words, considering the results along the $x$-axis as an example, when $\rho_{1x}$ is relatively large, the aggregation effect remains weak within the range of $\pm\chi$, allowing the diffusion effect to dominate and resulting in a tendency toward uniform density. However, outside this $\pm\chi$ range, the aggregation effect becomes stronger and overtakes the diffusion effect, leading to a steep density gradient near the boundary. Conversely, when $\rho_{1x}$ is relatively small, the aggregation effect is strong and dominant within $\pm\chi$, promoting density inhomogeneity. Outside this range, the aggregation effect weakens, allowing the diffusion effect to dominate and resulting in a more gradual density gradient near the boundary. From another perspective, within the definitions of $J_{Ax}, J_{Ay}$, the parameters $\rho_{1x}, \rho_{1y}$ can be interpreted as the degree of distance sensitivity in aggregation within the formed domain. As illustrated above, smaller values of $\rho_{1x}$ correspond to higher distance sensitivity in aggregation, whereas larger values indicate lower distance sensitivity.

Furthermore, by fixing $\rho_{1x} = 0.25$ and adopting $\rho_{1y} = 2.87$ - the value that maximizes the area S

(as detailed in the next section) - we compare the resulting analytical solution with the structure of the carapace in adult *M.japonica* (i.e., at the stage when the individual has reached full size).

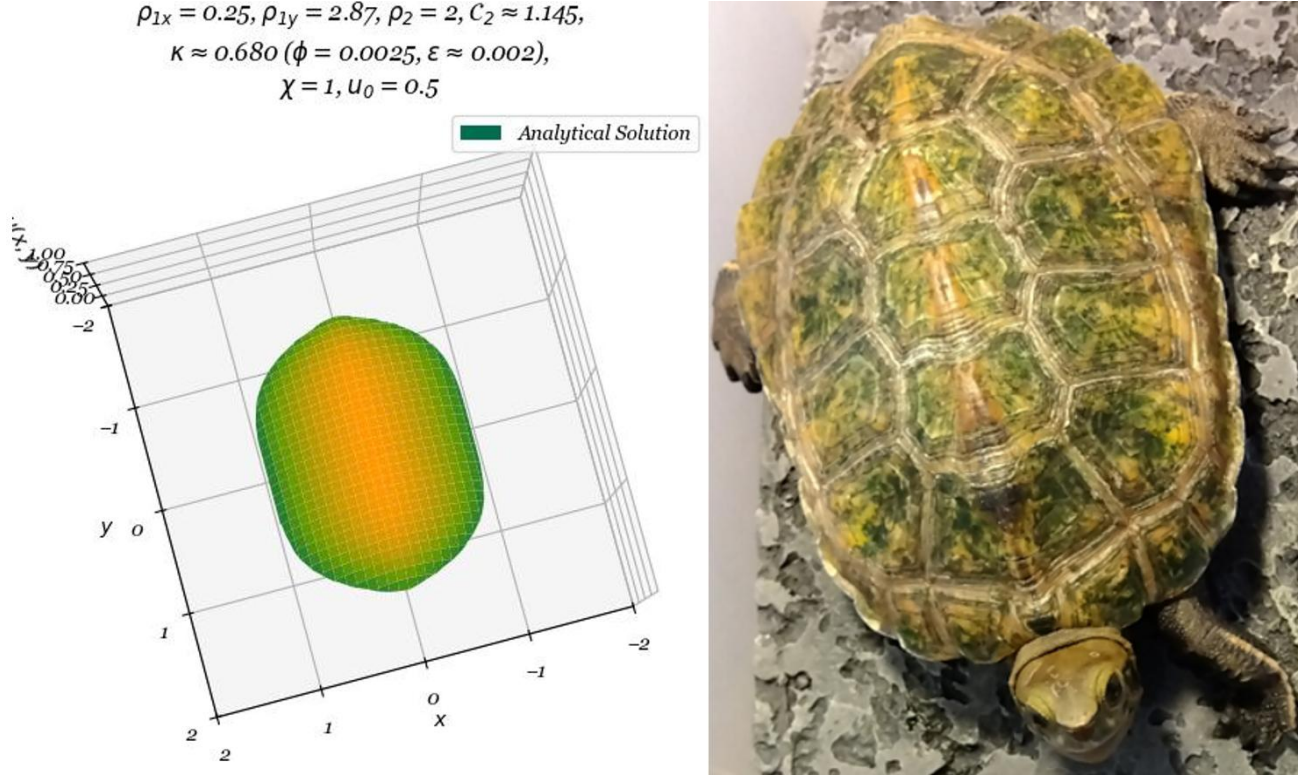

**Fig. 4**: Based on (7), the following parameters are set, and the resulting numerical analysis of $u(x, y)$ is presented in the graph on left: $\rho_{1x} = 0.25, \rho_{1y} = 2.87, \rho_2 = 2, \kappa \approx 0.680$ (based on $\phi = 0.0025, \varepsilon \approx 0.002$), $\chi = 1, u_0 = 0.5, l_2 = 3$, which leads to $C_2 \approx 1.145$. The darker regions represent higher values of $u$, while the lighter regions represent lower values. On the other hand, the image on the right shows *M.japonica*. A comparison between the carapace of *M.japonica* and the graph on the left reveals a similarity in both the overall domain structure and the midline ridge. (Figures in Color)

## 3.2. Distance Sensitivity of Aggregation and the Area of the Formed Domain

Furthermore, the area $S(\rho_{1x}, \rho_{1y})$, enclosed by the boundary defined as a function of the distance sensitivity of aggregation, takes the following form, and it is confirmed that it attains a maximum value at approximately $S \approx 4.97$, with $\rho_{1x} = 6.66$, and $\rho_{1y} = 6.66$. In particular, for the previously discussed case with $\rho_{1x} = 0.25$, indicated by the bold curve, the area reaches a maximum of approximately $S \approx 4.27$ at $\rho_{1y} = 2.87$.

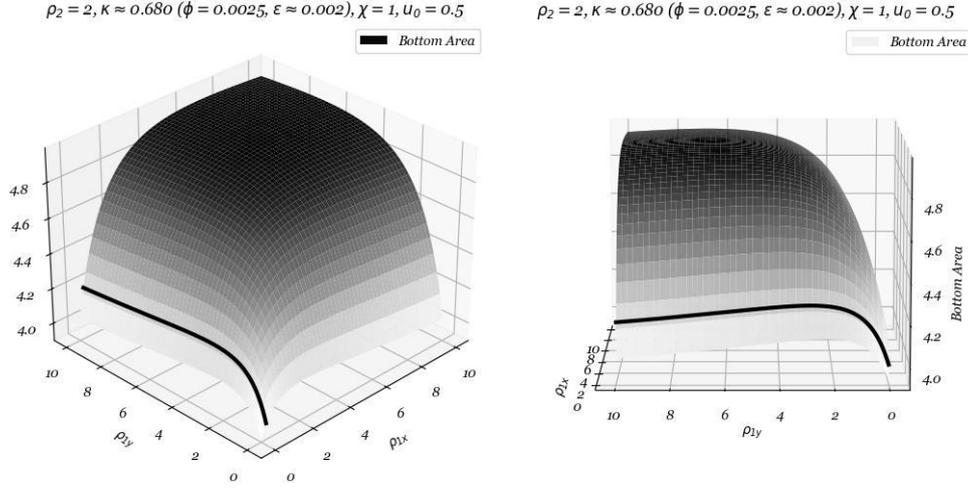

**Fig. 5**: Based on (9), the following parameters are set, and the resulting numerical analysis of $S(\rho_{1x}, \rho_{1y})$ is presented in the graph: $\rho_2 = 2, \kappa \approx 0.680$ (based on $\phi = 0.0025, \varepsilon \approx 0.002$), $\chi = 1, u_0 = 0.5, l_2 = 3$. The darker regions represent higher values of $u$, while the lighter regions represent lower values. To provide a comprehensive view of the area $S$, the graphs on the left and right are shown from different viewing angles; however, they represent the same graph, with only the viewing perspective altered. The bold curve indicates the area $S$ plotted under the condition where $\rho_{1x} = 0.25$ is fixed.

In other words, it can be interpreted that the lower the distance sensitivity of aggregation, the larger the resulting area $S$ tends to be.

### 3.3. Formation of Multiple High-Density Regions

Next, we numerically analyze the solution in cases where the aggregation flux density changes from positive to negative at two distinct points.

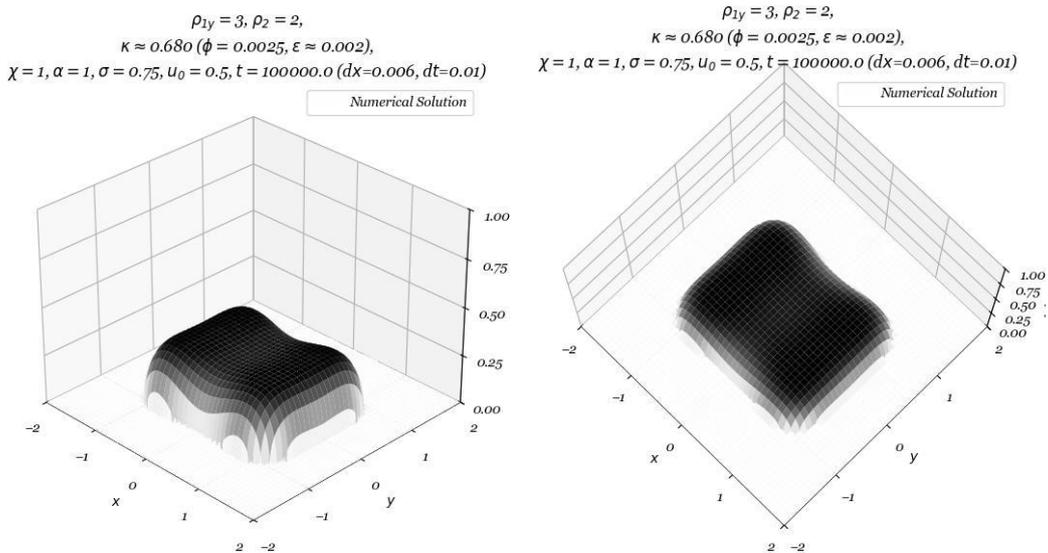

**Fig. 6**: In (6), only $J_{Ax}$, as defined in (10) as follows, is modified. Based on this, the following parameters are set, and the resulting numerical analysis of $u(t, x, y)$ is presented in the graph: $\rho_{1y} = 3, \rho_2 = 2, \kappa \approx 0.680$ (based on $\phi = 0.0025, \varepsilon \approx 0.002$), $\chi = 1, \alpha = 1, \sigma = 0.75, u_0 = 0.5, l_2 = 3, t = 100{,}000$ (dx = 0.006, dt = 0.01). The darker regions represent higher values of $u$, while the lighter regions represent lower values. The graph on the left is drawn from a horizontal perspective to show the uniformity or non-uniformity of the density $u$, while the graph on the right is drawn from a top-down view to show the boundaries of density $u$. These graphs are the same, with only the viewing angle changed.

As anticipated, the numerical solution exhibits localized high-density regions at two distinct locations. Furthermore, it is confirmed that the definition of the aggregation flux density also results in the formation of a well-defined domain (boundary).

## 4. Conclusion

As demonstrated in the previous section, when cells engage in (i) density-dependent diffusive movement and (ii) distance-dependent aggregative movement, localized regions of high density can emerge, and domain boundaries can form. Consistent with findings from prior studies, when the distance sensitivity of aggregation is low, the resulting cell density tends to become uniform. In contrast, when the distance sensitivity of aggregation is high, the aggregation effect outweighs the diffusive effect, leading to the emergence of spatial heterogeneity in density. Furthermore, if diffusion is interpreted as a repulsive force and aggregation as an attractive force, the observed steep density gradients at domain boundaries are also consistent with previous conclusions (Shigesada et al., 1979; Mogilner and Edelstein-Keshet, 1999). As previously noted, this distance sensitivity of aggregation can be understood as a potential for movement driven by haptotaxis, where cells migrate along gradients in the density of structural components such as ECM (Murray, 2013). In other words, the following component of the aggregation term can be interpreted as a potential governing aggregative movement:

$$-\varepsilon \left| \frac{x}{\chi} \right|^{\rho_{1x}} \text{sgn}(x), \text{or} -\varepsilon \left| \frac{y}{\chi} \right|^{\rho_{1y}} \text{sgn}(y). \qquad (11)$$

This conclusion aligns with the intuitive expectation that a domain boundary becomes defined when a balance is achieved between diffusive and aggregative movement, and that localized high-density regions may emerge depending on the strength of the aggregation effect. This analytical result may contribute to explaining the mechanisms underlying phenomena such as keel formation on the carapace of turtles and the determination of carapacial boundaries. Furthermore, both analytical and numerical investigations confirmed that the emergence of localized high-density regions occurs at points where the aggregation term changes sign from positive to negative. This fundamental principle also implies the potential for forming not just one or two, but multiple high-density regions (see Appendix D). We propose that such a simple aggregation–diffusion equation is capable of generating diverse density gradients and domains, and holds promise for advancing our understanding of turtle carapace development.

## 5. Discussion

This study, following previous research (Shigesada et al., 1979), adopts the parameter value $\rho_2 = 2$, but we also examine the solutions for other values, such as $\rho_2 = 0.5, 1$. By substituting $\rho_2 = 0.5, 1$ into the analytical solution for the two-dimensional domain, the following results are obtained:

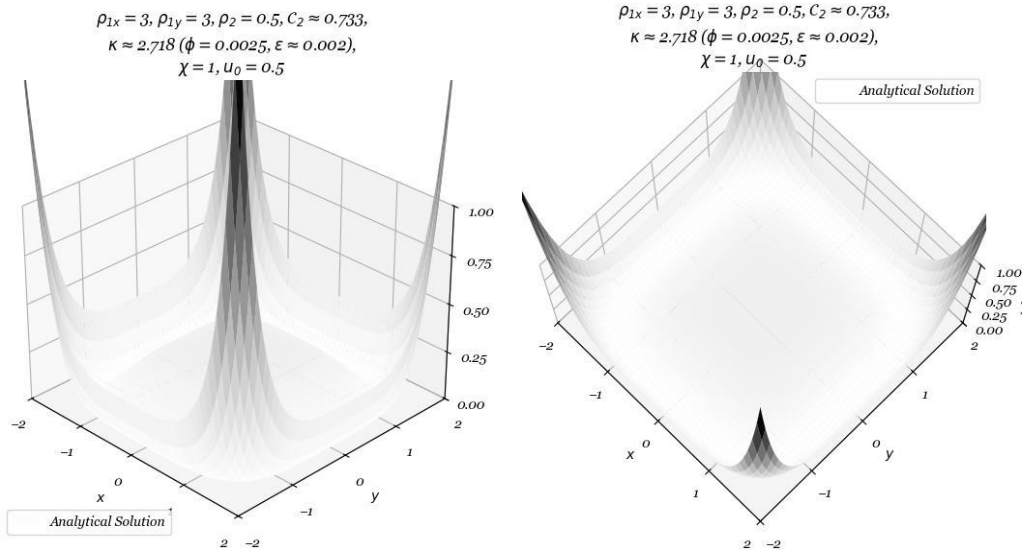

**Fig. 7**: Based on (7), the following parameters are set, and the resulting numerical analysis of $u(x, y)$ is presented in the graph: $\rho_{1x} = 3, \rho_{1y} = 3, \rho_2 = 0.5, \kappa \approx 0.680$ (based on $\phi = 0.0025, \varepsilon \approx 0.002$), $\chi = 1, u_0 = 0.5, l_2 = 3$, which leads to $\mathcal{C}_2 \approx 0.733$. The darker regions represent higher values of $u$, while the lighter regions represent lower values. The graph on the left is drawn from a horizontal perspective to show the uniformity or non-uniformity of the density $u$, while the graph on the right is drawn from a top-down view to show the boundaries of density $u$. These graphs are the same, with only the viewing angle changed.

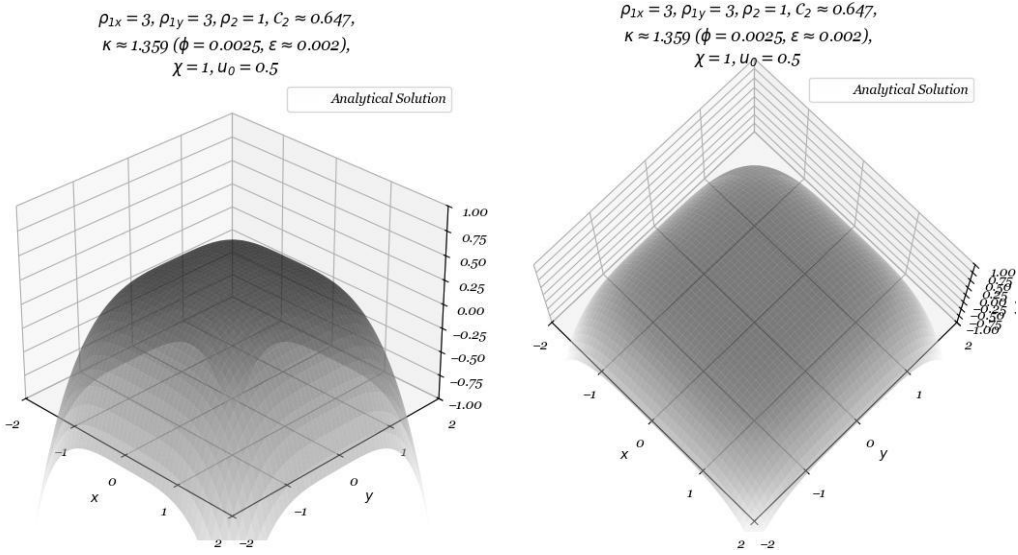

**Fig. 8**: Based on (7), the following parameters are set, and the resulting numerical analysis of $u(x, y)$ is presented in the graph: $\rho_{1x} = 3, \rho_{1y} = 3, \rho_2 = 1, \kappa \approx 0.680$ (based on $\phi = 0.0025, \varepsilon \approx 0.002$), $\chi = 1, u_0 = 0.5, l_2 = 3$, which leads to $C_2 \approx 0.647$. The darker regions represent higher values of $u$, while the lighter regions represent lower values. The graph on the left is drawn from a horizontal perspective to show the uniformity or non-uniformity of the density $u$, while the graph on the right is drawn from a top-down view to show the boundaries of density $u$. These graphs are the same, with only the viewing angle changed.

We define $u$ as the cell density, and therefore focus only on the regions where $u$ is positive. Furthermore, when $u = 0$,

$$J_{Ax} = J_{Ay} = 0, \text{and } J_D = 0 .$$  (12)

That is, in regions where $u = 0$, neither aggregation nor diffusion occurs, and these regions can be regarded as boundaries. Additionally, it should be noted that we assume $u \to 0$ as $x \to \infty$. Therefore, for $\rho_2 = 0.5, 1$, it can be inferred that a closed domain is formed, bounded by the regions where $u = 0$. In other words, similar results may be obtained for values of $\rho_2$ other than $2$ in real biological phenomena.

This study assumes that the cells capable of forming the keel undergo aggregation based on haptotaxis. We propose that, in the turtle carapace, tensile forces generated during neural tube closure (NTC) and the fusion of vertebral placodes may correspond to such aggregation movements (Cherepanov, 2006; Morita et al., 2012). In fact, during embryonic development in *M.reevesii* and *S.odoratus*, the keel along the midline and the two lateral keels appear morphologically distinct, possibly due to the former keel being formed through aggregation movements related to specific phenomena such as NTC and vertebral placode fusion. In particular, in the juvenile *M.reevesii* in Fig. 9, the two lateral keels clearly exhibit an arched shape, in contrast to the straight keel along the midline. Furthermore, previous studies have reported differences during embryonic development between vertebral and pleural placodes, as well as between pleural and marginal placodes (the former located above the septal invagination, and the latter within it) (Cherepanov, 2006). Assuming the existence of two distinct mechanisms for keel formation, a remaining question is: "By what mechanism are the two lateral keels formed?" We focus on the fact that these two lateral keels appear during rib growth, prior to the formation of the costal bones (Paredes et al., 2020). FGF10 secreted from mesenchymal tissue in CR induces the secretion of FGF8 at the distal tips of the ribs, and FGF10 and FGF8 form a positive feedback loop that drives rib growth (Gilbert et al., 2001; Cebra-Thomas et al., 2005; Nagashima et al., 2007; Moustakas, 2008; Kaplinsky et al., 2013; Burke, 2015). Therefore, considering that the lateral keels appear during rib growth, it is possible that the CR and ribs play a role in their expression. Indeed, FGF10 is one of the growth factors involved in the development of keratinocytes (Zhao et al., 2024), which directly contribute to scute formation, and the regulatory mechanism of the FGF10-FGF8 positive feedback loop remains unresolved. These raise the possibility that a localized increase in FGF10 concentration along the CR might indirectly lead to the formation of the two lateral keels. Moreover, the CR is not only crucial for carapace formation, but the balance between FGF and BMP signaling pathways has also been reported to be essential (Heisenberg et al., 2000; Minina et al., 2002; Cebra-Thomas et al., 2005; Moustakas, 2008; Burke, 2015; Yang et al., 2022). In summary, we propose that both the CR and FGF signaling play vital roles in carapace formation, and may also contribute to the development of the two lateral keels observed on the carapace. Additionally, in the case of *M.japonica*, the two lateral keels appear to disappear in adulthood, while the single keel along the midline remains clearly visible in Fig. 9.

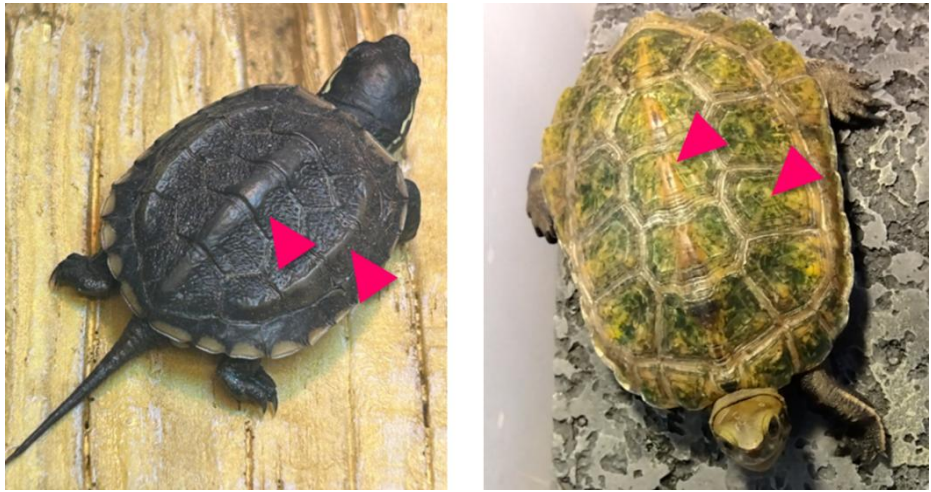

**Fig. 9**: The image on the left shows a juvenile *M.reevesii*, in which a single ridge along the midline of the carapace, as indicated by the triangular marker, is clearly visible. In addition, two curved lateral ridges can also be observed. In contrast, the image on the right shows an adult *M.japonica*, in which a single ridge along the midline is similarly observed, as indicated by the triangular marker, but the two curved lateral ridges are absent. (Figures in Color)

Based on the above observations, we consider it plausible to hypothesize the existence of two distinct mechanisms underlying keel formation. Accordingly, we hypothesize that the formation of the single keel along the midline of the carapace and that of the two lateral keels are governed by different mechanisms.

Moreover, previous studies (Baena-Lopez et al., 2005; Silva and Vincent, 2007) have reported the existence of oriented cell migration or division, in which the sensitivity to aggregation distance is higher in the lateral direction (*e.g.*, along the $x$-axis) and lower in the longitudinal direction (*e.g.*, along the $y$-axis), meaning that the diffusive effect is relatively greater in the latter. Such anisotropic directionality between the longitudinal and lateral axes may contribute to the emergence of keels as the carapace elongates. This notion may be supported by the patterns shown in Fig. 2 and Fig. 3. In the present study, the obtained analytical solutions assume isotropic diffusion; thus, the parameters $\rho_{1x}, \rho_{1y}$, representing the distance sensitivity of aggregation, can be interpreted as describing this directional bias.

On the other hand, it is also conceivable that keels may form as a result of the mechanisms involved in individual scute formation. For instance, if scutes exhibiting protrusions along the midline in the longitudinal direction are arranged in sequence, it may appear as though a single keel has formed along the midline of the carapace. However, previous studies (Alibardi, 2006; Moustakas-Verho et al., 2014) on scute formation have not reported any such mechanism of keel formation. Moreover, if such localized protrusions were to arise on individual scutes, one might expect similar protrusions to appear uniformly across all scutes. This raises the possibility that distinct mechanisms operate in different scute regions, which we consider to be an important subject for future research.

## 6. Author contributions

**Shin Nishihara:** Conceptualization, Methodology, Software, Validation, Formal analysis, Investigation, Resources, Data Curation, Writing - Original Draft, Writing - Review and Editing, and Visualization

**Toru Ohira:** Resources, Writing - Review and Editing, Supervision, Project administration, and Funding acquisition

## 7. Acknowledgements and Funding

This work was supported by JSPS Topic-Setting Program to Advance Cutting-Edge Humanities and Social Sciences Research Grant Number JPJS00122674991, and by Ohagi Hospital, Hashimoto, Wakayama, Japan.

## 8. Statements and Declarations

The authors declare that they have no known competing financial interests or personal relationships that could have appeared to influence the work reported in this paper.

## Appendix

### A.  Derivation of the analytical solution  $u(x)$

For sufficiently large values of $t$, the analytical solution $u(x)$ can be obtained by solving the following equation:

$$\varepsilon \left| \frac{x}{\chi} \right|^{\rho_1} u \, \mathrm{sgn}(x) + \phi \left( \frac{u}{u_0} \right)^{\rho_2} \frac{\partial u}{\partial x} = 0 \, . \tag{A1}$$

By applying the method of separation of variables and introducing a constant  $\mathcal{C}$, the following analytical solution  $u(x)$  is obtained:

$$\left( \frac{u}{u_0} \right)^{\rho_2} = -\frac{1}{\kappa \chi^{\rho_1}} \left( \frac{|x|^{\rho_1+1}}{\rho_1 + 1} \right) + \mathcal{C} \, . \tag{A2}$$

Next, to determine the constant  $\mathcal{C}$, we use the following two assumptions: (i) the total density  $2I$  is conserved over time, and (ii) the solution is symmetric with respect to the origin. That is, we solve the following equation:

$$I = \int_0^{x_{\text{boundary}}} \frac{u(x)}{u_0} \, \mathrm{d}x \, , \text{ where } x_{boundary} > 0 \, . \tag{A3}$$

The following variable transformation is applied:

$$\mathcal{C}\xi := \mathcal{C} - \frac{x^{\rho_1+1}}{\kappa(\rho_1 + 1)\chi^{\rho_1}} \tag{A4}$$

This variable transformation converts the original equation into the following form:

$$I = \left( \frac{\kappa \chi^{\rho_1}}{(\kappa(\rho_1 + 1)\chi^{\rho_1})^{\frac{\rho_1}{\rho_1+1}}} \right) \cdot \left( \mathcal{C}^{\frac{1}{\rho_2}+\frac{1}{\rho_1+1}} \right) \cdot \int_0^1 \xi^{\frac{1}{\rho_2}} (1 - \xi)^{-\frac{\rho_1}{\rho_1+1}} \, \mathrm{d}\xi \, . \tag{A5}$$

This allows us to determine the constant  $\mathcal{C}$:

$$\mathcal{C} = \left( \frac{\kappa \chi^{\rho_1}}{(\rho_1 + 1)^{\rho_1}} \right)^{-\frac{\rho_2}{\rho_1+\rho_2+1}} \cdot \left( \frac{I}{\mathrm{B} \left( \frac{1}{\rho_2} + 1, \frac{1}{\rho_1 + 1} \right)} \right)^{\frac{(\rho_1+1)\rho_2}{\rho_1+\rho_2+1}} \, . \tag{A6}$$

### B.  Derivation of the analytical solution  $u(x, y)$

As with the analytical solution  $u(x)$  on the one-dimensional domain, the constant  $\mathcal{C}_2$  is determined using the conserved quantity  $I_2$. Specifically, we solve the following equation:

$$I_2 = 4 \int_0^{x_C} \int_0^{y_x} \left( \mathcal{C}_2 - \frac{x^{\rho_{1x}+1}}{K_x} - \frac{y^{\rho_{1y}+1}}{K_y} \right)^{\frac{1}{\rho_2}} \mathrm{d}y \, \mathrm{d}x \, . \tag{B1}$$

In this equation, symmetry with respect to the origin is applied. First, we perform the following

transformation:

$$\mathcal{C}_2' := \mathcal{C}_2 - \frac{x^{\rho_{1x}+1}}{K_x} \text{ , and } \mathcal{C}_2'\eta := \mathcal{C}_2' - \frac{y^{\rho_{1y}+1}}{K_y} \ . \tag{B2}$$

Applying this transformation yields the following equation:

$$I_2 = \frac{4K_y^{\frac{1}{\rho_{1y}+1}}}{\rho_{1y}+1} \int_0^{x_C} \mathcal{C}_2'^{\frac{1}{\rho_{1y}+1}+\frac{1}{\rho_2}} \int_0^1 \eta^{\frac{1}{\rho_2}} (1-\eta)^{-\frac{\rho_{1y}}{\rho_{1y}+1}} \mathrm{d}\eta \, \mathrm{d}x,$$

$$\Rightarrow I_2 = \frac{4K_y^{\frac{1}{\rho_{1y}+1}}}{\rho_{1y}+1} \cdot \mathrm{B}\left(\frac{1}{\rho_2}+1, \frac{1}{\rho_{1y}+1}\right) \cdot \int_0^{x_C} \left(\mathcal{C}_2 - \frac{x^{\rho_{1x}+1}}{K_x}\right)^{\frac{1}{\rho_{1y}+1}+\frac{1}{\rho_2}} \mathrm{d}x \ . \tag{B3}$$

Furthermore, the following transformation is applied:

$$\mathcal{C}_2\xi := \mathcal{C}_2 - \frac{x^{\rho_{1x}+1}}{K_x} \ . \tag{B4}$$

This variable transformation yields the following constant $\mathcal{C}_2$:

$$I_2 = \mathcal{C}_2^{\frac{1}{\rho_{1x}+1}+\frac{1}{\rho_{1y}+1}+\frac{1}{\rho_2}} \cdot \frac{4K_x^{\frac{1}{\rho_{1x}+1}}K_y^{\frac{1}{\rho_{1y}+1}}}{(\rho_{1x}+1)(\rho_{1y}+1)} \cdot \mathrm{B}\left(\frac{1}{\rho_2}+1, \frac{1}{\rho_{1y}+1}\right) \cdot \mathrm{B}\left(\frac{1}{\rho_{1y}+1}+\frac{1}{\rho_2}+1, \frac{1}{\rho_{1x}+1}\right) \ . \tag{B5}$$

## C. Derivation of the area of the domain $S$

By using symmetry with respect to the origin, the area enclosed by the boundary is described as follows:

$$S(\rho_{1x}, \rho_{1y}) = 4K_y^{\frac{1}{\rho_{1y}+1}} \int_0^{x_{boundary}} \left(\mathcal{C}_2 - \frac{x^{\rho_{1x}+1}}{K_x}\right)^{\frac{1}{\rho_{1y}+1}} \mathrm{d}x \ . \tag{C1}$$

The following variable transformation is performed:

$$\mathcal{C}_2\zeta := \mathcal{C}_2 - \frac{x^{\rho_{1x}+1}}{K_x} \ . \tag{C2}$$

Through this variable transformation, the area $S$ is obtained as follows:

$$S(\rho_{1x}, \rho_{1y}) = \frac{4K_x^{\frac{1}{\rho_{1x}+1}}K_y^{\frac{1}{\rho_{1y}+1}}\mathcal{C}_2^{\frac{1}{\rho_{1x}+1}+\frac{1}{\rho_{1y}+1}}}{\rho_{1x}+1} \int_0^1 \zeta^{\frac{1}{\rho_{1y}+1}} (1-\zeta)^{-\frac{\rho_{1x}}{\rho_{1x}+1}} \mathrm{d}\zeta \ . \tag{C3}$$

## D. Expression of Three High-Density Regions

By applying the fundamental principle that the expression of localized high-density regions occurs at the point where the value in the aggregation term changes from positive to negative, it is possible, for example, to form three high-density regions:

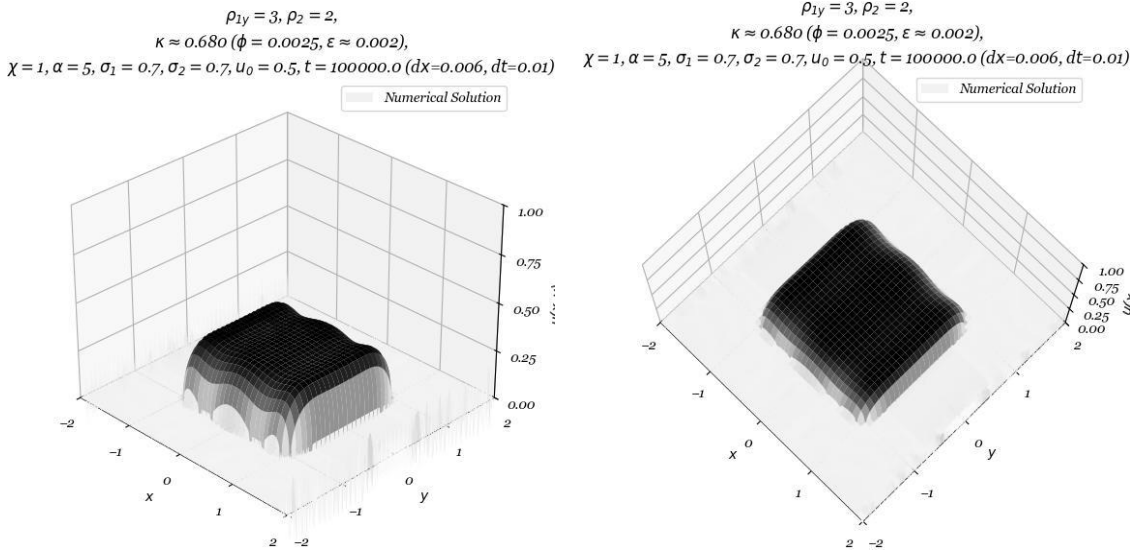

**Fig. D1**: In (6), only $J_{Ax}$, as defined in (D1) as follows, is modified. Based on this, the following parameters are set, and the resulting numerical analysis of $u(t, x, y)$ is presented in the graph: $\rho_{1y} = 3, \rho_2 = 2, \kappa \approx 0.680$ (based on $\phi = 0.0025, \varepsilon \approx 0.002$), $\chi = 1, \alpha = 5, \sigma_1 = 0.7, \sigma_2 = 0.7, u_0 = 0.5, l_2 = 3, t = 100{,}000$ (dx = 0.006, dt = 0.01). The darker regions represent higher values of $u$, while the lighter regions represent lower values. The graph on the left is drawn from a horizontal perspective to show the uniformity or non-uniformity of the density $u$, while the graph on the right is drawn from a top-down view to show the boundaries of density $u$. These graphs are the same, with only the viewing angle changed.

$$J_{Ax} := -\varepsilon\alpha\left(\frac{x}{\chi}\right)\left(\left(\frac{x}{\chi}\right)^2 - \sigma_1^2\right)\left(\left(\frac{x}{\chi}\right)^2 - \sigma_2^2\right)u \text{ , where } \alpha, \sigma_1 \text{ and } \sigma_2 \text{ are positive constants,} \qquad (D1)$$